\documentclass[pra,twocolumn,aps,showpacs,superscriptaddress]{revtex4-1}

\usepackage{amsmath}
\usepackage{amssymb}
\usepackage{hyperref}
\usepackage{graphicx}
\usepackage{subfigure}
\usepackage[usenames,dvipsnames]{color}

\begin{document}


\title{Formation, dynamics and stability of coreless vortex dipoles in 
       phase-separated binary condensates}

\author{S. Gautam} 
\affiliation{Physical Research Laboratory,
             Navarangpura, Ahmedabad - 380 009, India}
\author{P. Muruganandam} 
\affiliation{ School of Physics, 
              Bharathidasan University, Tiruchirapalli 620 024, 
              Tamil Nadu, India}
\author{D. Angom}
\affiliation{Physical Research Laboratory,
         Navarangpura, Ahmedabad - 380 009, India}


\date{\today}
\begin{abstract}
 We study the motion of the Gaussian obstacle potential created by blue 
detuned laser beam through a phase-separated binary condensate in 
pancake-shaped traps. For the velocity of the obstacle above a critical 
velocity, we observe the generation of vortex dipoles in the outer component 
which can penetrate the inner component. This is equivalent to finite, 
although small, transport of outer component across the inner component. In 
the inner component, the same method can lead to the formation of coreless 
vortex dipoles.  
\end{abstract}


\maketitle


\section{Introduction}

Generation of vortex dipoles in dilute Bose-Einstein condensates (BECs) when a 
superfluid moves past an obstacle with a velocity greater than a critical 
velocity (speed of sound locally) has been established by various numerical 
studies \cite{Frisch}-\cite{Raman}. It has been established, both 
theoretically and numerically, that even the flow of the BEC below bulk sound 
speed can result in the flow velocity greater than the local sound speed at 
the equator of the obstacle. It implies that the critical velocity for the 
creation of vortex-antivortex pairs is lower than the bulk sound speed. Above 
this critical speed vortex-antivortex pairs are created periodically, 
resulting in an oscillatory drag force. This is due to the phase difference 
between the main flow and stationary wake behind the obstacle. Whenever this 
phase difference grows to $2\pi$ a vortex-antivortex pair is created. 
Recently, Neely et al. \cite{Neely} observed the generation of vortex dipoles 
in a pancake-shaped $^{87}$Rb condensate, when the condensate was moved past 
a Gaussian obstacle, created by blue-detuned laser beam, above critical 
velocity. Real time dynamics of single vortex line and vortex dipoles has 
also been studied experimentally \cite{Freilich}. The flow of the miscible 
binary and spinor condensates across a Gaussian obstacle potential has also 
been investigated theoretically in Refs.\cite{Susanto,Gladush,Rodrigues}.
Vortex dipoles or vortex rings are also produced as one of the the decay 
products of dark solitons (a notch or density depression in the condensate 
with a phase slip of $\pi$ across it), due to the onset of a transverse 
instability in 2D and 3D geometries, namely snake instability. The decay of 
dark soliton into vortex rings has already been observed experimentally 
\cite{Anderson, Dutton}. In this context, it should be noted that motion of 
the obstacle across a 1D BEC generates grey solitons 
\cite{Hakim,Radouani,Carretero}. The snake instability is also responsible for 
the decay of ring soliton into a necklace of vortex dipoles \cite{Theocharis}.

In the present work, we study the flow of phase-separated binary condensates 
across a Gaussian obstacle numerically. We consider two experimentally 
realizable binary condensates: the first of these has $^{85}$Rb and $^{87}$Rb
and the second has the two hyperfine states of $^{87}$Rb as the two 
constituent species. In both these condensates, one of the scattering length
can be tuned by using magnetic Feshbach resonances \cite{Papp, Tojo}.
This allows us to choose the tunable scattering lengths values suitable for 
generation of coreless vortex dipoles. We observe the generation of vortex 
dipoles for the flow velocity greater than the critical velocity. In our 
simulations while moving the obstacle, its strength is continuously decreased. 
We consider two cases: (a) the obstacle laser potential, initially in the 
outer component, is moved towards the center, and (b) the obstacle laser 
potential, initially in the inner component, moves towards the interface of
the binary condensate. With the successful realization of vortex dipoles in 
single component condensates using obstacle potential \cite{Neely}, the 
present studies give more than enough reasons to extend these experimental 
studies to binary condensates.


\section{Binary condensates and vortex dipoles}

\subsection{Meanfield description of binary condensates}

The dynamics of binary condensate at $T=0$K can be very well described by
a set of coupled GP equations
\begin{equation}
 \left[ \frac{-\hbar^2}{2m}\nabla^2 + V_i({\mathbf r},t) + 
 \sum_{j=1}^2U_{ij}|\Psi_j({\mathbf r},t)|^2 - 
 i\hbar\frac{\partial}{\partial t}\right]\Psi_i ({\mathbf r},t) = 0
 \label{eq.gp1}
\end{equation}
in mean field approximation, where $i = 1, 2$ is the species index. Here 
$U_{ii} = 4\pi\hbar^2a_{ii}/m_i$, where $m_i$ is the mass and $a_{ii}$ is 
the $s$-wave scattering length, is the intra-species interaction, 
$U_{ij}=2\pi\hbar^2a_{ij}/m_{ij}$, where $m_{ij}=m_i m_j/(m_i+m_j)$ is the
reduced mass and $a_{ij}$ is the inter-species scattering length, is the
inter-species interaction, and $V_i({\mathbf r})$ is the trapping potential 
experienced by $i$th species. In the present work, we consider binary 
condensate consisting either of $^{85}$Rb and $^{87}$Rb for which 
$m_1\approx m_2$. Furthermore, we also consider identical trapping potential 
for both the species, the total potential is then $i$th species 
\begin{equation}
 V({\mathbf r},t) = \frac{m_1\omega^2}{2}(x^2 + \alpha ^2 y^2 + 
                      \beta ^2 z^2) + V_{\rm obs}(x,y,t),
\label{eq.pots}
\end{equation}
where $V_{\rm obs}(x,y,t) = 
V_0 (t)\exp\lbrace -2([x-x_0(t)]^2+y^2)/w_0^2\rbrace$ is the potential 
created by blue detuned laser beam, and $\alpha$ and $\beta$ are anisotropy 
parameters. To rewrite the coupled GP equations in scaled units, define the 
oscillator length of the trapping potential
\begin{equation}
   a_{\rm osc} = \sqrt{\frac{\hbar}{m\omega}},
\end{equation}
and consider $\hbar\omega$ as the unit of energy. We then divide the 
Eq.(\ref{eq.gp1}) by $\hbar\omega$ and apply the transformations
\begin{equation}
\tilde{{\mathbf r}}   =  \frac{{\mathbf r}}{a_{\rm osc}}, 
~\tilde{t} =  t\omega, \text{and} ~
\phi_{i}(\tilde{{\mathbf r}},\tilde{t})=   \sqrt{\frac{a_{\rm osc}^3}{N_i}}
\Psi_i({\mathbf r},t).
\end{equation}
The transformed coupled GP equation in scaled units is
\begin{equation}
  \left[ -\frac{\tilde{\nabla}^2}{2} + V_i(\tilde{{\mathbf r}},\tilde{t}) + 
  \sum_{j=1}^2\tilde{U}_{ij}|\phi_j(\tilde{{\mathbf r}},\tilde{t})|^2 - 
   i\frac{\partial}{\partial \tilde{t}}\right]
  \phi_i (\tilde{{\mathbf r}},\tilde{t}) = 0,
\label{eq.gp2}
\end{equation}
where $\tilde{U}_{ii} = 4\pi a_{ii}N_i/a_{\rm osc}$, 
$\tilde{U}_{ij} = 4\pi a_{ij}N_j/a_{\rm osc}$, and 
$\tilde{\nabla}^2 = a_{\rm osc}^2\nabla^2$. From here on, we will represent the 
scaled quantities without tilde unless stated otherwise.

   In pancake-shaped (highly oblate) traps, trapping frequency along axial 
direction is much greater than the radial directions, 
i.e., $\beta\gg\alpha\sim1$; in which case Eq.(\ref{eq.gp2}) can be reduced 
to two dimensional form by substituting $\phi(\mathbf{r},t)= 
\psi(x,y,t)\xi(z)\exp({-i\beta t/2})$ \cite{Muruganandam}. Here 
$\xi=(\beta/(2\pi))^{1/4}\exp(-\beta z^2/4)$, the ground state wave function 
in axial direction. The reduced two dimensional form of coupled GP equation in
pan-cake shaped traps is
\begin{eqnarray}
  \left[ -\frac{1}{2}\left(\frac{\partial^2}{\partial x^2} + 
  \frac{\partial^2}{\partial y^2}\right) + \frac{x^2+\alpha_i^2y^2}{2} + 
  U(x,y,t) + \right.\\
 \sum_{j=1}^2 u_{ij}|\psi_j({\mathbf r},t)|^2 \left. - 
  i\frac{\partial}{\partial t}\right]
  \psi_i ({\mathbf r},t) = 0,\nonumber
\label{eq.gp3}
\end{eqnarray}
where $ u_{ii} = 2 a_{ii}N_i\sqrt{2\pi\beta_i}/a_{\rm osc}$ and 
$ u_{ij} = 2 a_{ij}N_j\sqrt{2\pi\beta_i}/a_{osc}$. In the present work, we
consider $u_{12}>\sqrt{u_{11}u_{22}}$ so that the ground state of the binary 
condensate is phase-separated. The ground states of the binary condensates
in the phase separated domain have been studied in Refs.\cite{Ho}-\cite{Gautam2} 
In pancake-shaped traps the binary condensates has cylindrical interface 
separating the two components \cite{Gautam1}.


\subsection{Vortex dipole trajectory in a single component condensate}
   The total velocity field experienced by each vortex in vortex dipole is the
vector sum of two component fields. One of these fields is the velocity field
generated on each vortex due to inhomogeneous nature of the condensate in 
the trapping potential. This inhomogeneity generated velocity field is 
responsible for the rotation of an off center vortex around the trap center
\cite{Jackson-2,Svidzinsky}. In addition to this, each vortex creates a 
velocity field varying inversely with the distance from its center, which
is experienced by the other vortex of vortex dipole. If $(x,y)$ and $(x,-y)$ 
are the locations of the positively and negatively charged vortices of the 
vortex dipole, respectively, then the total velocity field experienced by the 
positively charged vortex of the vortex dipole  
\begin{equation}
 \mathbf v(x,y) = \omega_{\rm pr}\hat k \times \mathbf r + \frac{1}{2y}\hat i, 
\end{equation}
where $\omega_{\rm pr}$ is the rotational frequency of a vortex with charge 
$+1$ in the condensate. In terms of component velocities, the previous equation
can be written as
\begin{eqnarray}
 \frac{dx}{dt} & = & -\omega_{\rm pr}y + \frac{1}{2y}, \nonumber\\
 \frac{dy}{dt} & = & \omega_{\rm pr}x.
 \label{vdipole_trejectory1} 
\end{eqnarray}
This coupled set of differential equations describes the trajectory of the
positively charged vortex of the vortex dipole. In the present work, we move
the obstacle along $x$-axis, leading to the generation of vortex dipole located
symmetrically about $x$-axis. Since $y$ is small (and positive) for the 
positively charged vortex of the vortex dipole at the instant of generation, 
one can neglect $-\omega_{\rm pr}y$ in first equation of coupled 
Eqs.\ref{vdipole_trejectory1} to obtain
\begin{eqnarray}
 \frac{dx}{dt} & \approx & \frac{1}{2y}, \nonumber\\
 \frac{dy}{dt} & = & \omega_{\rm pr}x.
 \label{vdipole_trejectory2}
\end{eqnarray}
Eqs.\ref{vdipole_trejectory2} imply that vortex initially to the left
of the origin moves towards origin with increasing speed along
positive $x$-axis but decreasing speed along negative $y$-axis.
At $x=0$, $dy/dt=0$ and beyond this vortex moves away from origin
with decreasing speed along $x$-axis but increasing speed along positive 
$y$-axis. Near the origin $\omega_{\rm pr}x\ll1/2y$, then the velocity of 
the vortex can be approximated as
\begin{equation}
 \mathbf v(x,y) \approx \frac{1}{2y}\hat i.
\end{equation}
In this case the velocity of the vortex is mainly induced one, and it moves
with induced velocity towards positive $x$ direction. On the other hand if
$y$ is too large, one can neglect $1/2y$ term in Eq.\ref{vdipole_trejectory1}
to obtain
\begin{equation}
 \mathbf v(x,y) \approx \omega_{\rm pr}\hat k \times \mathbf r.
\end{equation}
In this case, the velocity of the vortex is mainly due to the inhomogeneous
nature of the condensate, resulting in the rotation with frequency 
$ \omega_{\rm pr}$. The transition between vortex induced velocity to 
inhomogeneity induced velocity occurs when $dx/dt = 0$, i.e., velocity
along x-axis changes sign. From Eqs.\ref{vdipole_trejectory1}, it occurs
at $y = \sqrt{1/2\omega_{\rm pr}}$.  


\section{Obstacle modified condensate density}

 In the phase separated domain of binary condensates, the species with lower
repulsion energy forms a core  and the other species forms a shell around it.
For convenience, we identify the former and later as the first and second 
species, respectively.  With this labelling, interaction energies 
$u_{11}<u_{22} $ and for equal populations this implies $a_{11}<a_{22}$. There 
is large difference of the local coherence length $\xi_i=1/\sqrt{2n_iu_{ii}}$ 
when $u_{11}\ll u_{22}$. This results in different density modifications
for the two species when an obstacle beam is applied. The dynamics is most 
sensitive to the coherence length difference while the obstacle is crossing
the interface layer of the binary condensate.

\begin{figure}[ht]
\includegraphics[width=8.5cm] {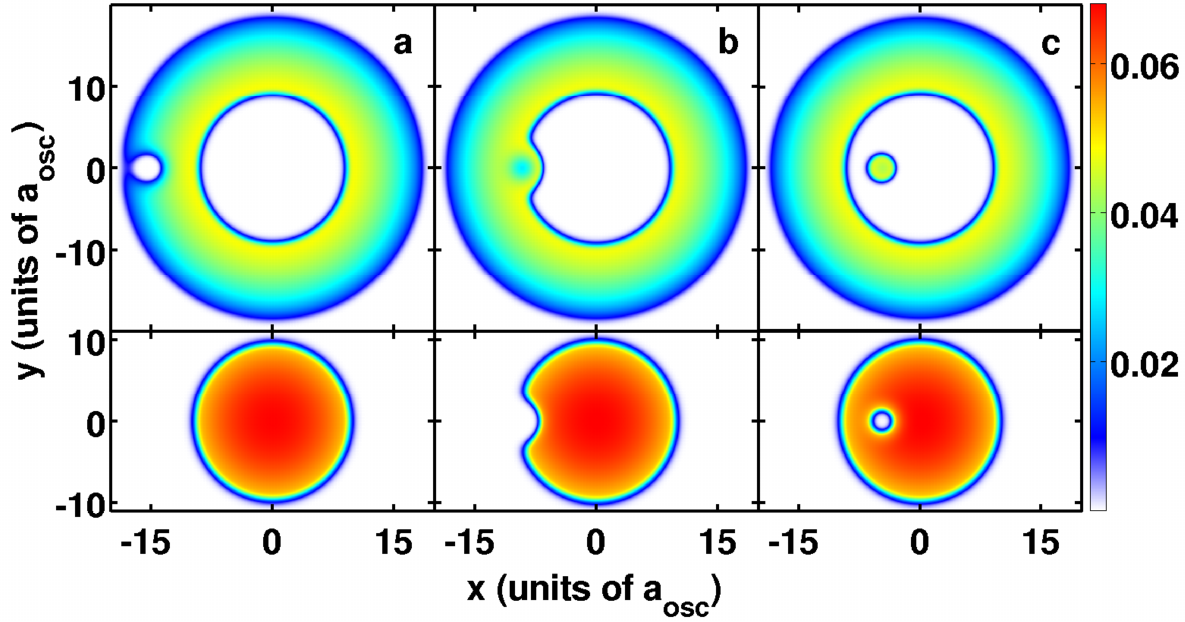}
\caption{The stationary state wave functions of $^{85}$Rb-$^{87}$Rb binary
condensate obtained from imaginary time propagation. For (a)
the obstacle is located near $R_{\rm out}$, for (b) the obstacle is located at
the interface, and for (c) it is well inside the bulk of the inner species.}
\label{plot2}
\end{figure}

  To analyze the density perturbations in presence of the obstacle beam, 
let $R_{\rm in}$ be the radius of the inner species or the interface boundary. 
And, let $R_{\rm out}$ be the radial extent of the outer species. In the 
absence of the obstacle beam, the chemical potential of first species is 
$\mu_1=R_{\rm in}^2/4 + u_{11}/(\pi  R_{\rm in}^2)$. Here, the second term is 
correction arising from the presence of the second species. The chemical
potential of the second species is $R_{\rm out}^2/2$. Both of these are in
scaled units. We consider the obstacle beam initially ($t=0$) located 
at $(-R_{\rm out},0)$ and traversing towards the center with velocity 
$v_{\rm obs}$. As it moves, its intensity is ramped down at the rate $\eta$. 
The location of the beam at a later time is
\begin{equation}
   x_0(t) = -R_{\rm out} + v_{\rm ob}t,
\end{equation}
and  intensity of the beam is
\begin{equation}
  V_0(t) = V_0(0) -\eta t,
\end{equation} 
where  $V_0(0)$ is the initial intensity of the obstacle beam. At the starting 
point, the total potential $V(R_{\rm out},0,0) > R_{\rm out}^2/2$ and the 
density of the outer species $|\psi_2|^2$ is zero around the center of the 
obstacle beam as is shown in Fig. \ref{plot2}(a).
However, as it traverses the 
condensates with decreasing intensity, at some later time 
$t'$, $V(x_0(t'),0,0) < R_{\rm out}^2/2$. Density $|\psi_2|^2$ is then finite
within the obstacle. For compact notations, hereafter we drop the explicit 
notation of time dependence while writing $x_0(t)$ and  $V_0(t)$.

\subsection{Density modification at interface}

  As the obstacle beam approaches the interface, the intensity is weak to 
expel out the outer species. However, when the beam enters the inner species
with any finite intensity $V(R_{\rm in}, 0,t) > \mu_1 $, it is sufficient to 
make $|\psi_1|^2$ zero around the beam [see Fig.~\ref{plot2}(a)]. The interface
is modified and equipotential curve of the interface is
\begin{equation}
  \frac{x^2+y^2}{2} + V_0 \exp\left[ -2\frac{(x-x_0)^2+y^2}{w^2}\right] 
      = \frac{R_{\rm in}^2}{2} .
\label{eq_pot_surface}
\end{equation}
Around the center of the obstacle beam, $\sqrt{(x-x_0)^2+y^2}<w/\sqrt{2}$, 
and the obstacle potential is 
\begin{equation}
  V_{\rm opt}\approx V_0-\frac{2V_0}{w^2}\left [ (x-x_0)^2+y^2\right].
 \label{app_obs_pot}
\end{equation}
Hence from Eq.(\ref{eq_pot_surface}) the equipotential curve in the 
neighborhood of the obstacle beam is
\begin{equation}
 \left(x+\frac{C}{2A}\right)^2 + y^2 = \tilde{R}_{\rm in}^2,
\end{equation}
where 
\begin{eqnarray}
    A  & = &\frac{1}{2}-\frac{2V_0}{w^2}, \nonumber \\
    C  & = & \frac{4V_0x_0}{w^2}, \nonumber  \\
    \tilde{R}_{\rm in}^2 & = & \frac{1}{A}\left (\frac{R_{\rm in}^2}{2} 
       +\frac{C^2}{4A} + C\frac{x_0}{2} - V_0\right ).    \nonumber
\end{eqnarray}
Equipotential curve in the vicinity of the obstacle is thus modified into a 
circle with center at ($-C/2A,0$) and radius $\tilde{R}_{\rm in}$. In our 
numerical calculations, the Gaussian obstacle potential is quite narrow and
$w^2/(4V_0)\ll1$. Retaining only the leading order terms in $w^2/4V_0$, 
the center is approximately located at $(x_0+w^2x_0/(4V_0), 0)$ and radius is
\begin{equation}
   \tilde{R}_{\rm in} \approx\sqrt{\frac{w^2}{2V_0}\left(1
   +\frac{w^2}{4V_0}\right)\left(V_0
              +\frac{x_0^2}{2}-\frac{R_{\rm in}^2}{2}\right)}.
\end{equation}
Sufficiently away from the obstacle $\sqrt{(x-x_0)^2+y^2}> w/\sqrt{2}$ and the 
obstacle potential is negligible. The equipotential curve then remains 
unchanged, $R_{\rm in}^2=x^2+y^2$. Thus an approximate piecewise expression of 
the equipotential curve is
\begin{eqnarray}
  \left(x+\frac{C}{2A}\right)^2+y^2 &=& \tilde{R}_{\rm in}^2,
           \text{ if }  |R_{\rm in}-x_0|<\frac{w}{\sqrt{2}}, \nonumber \\
   x^2+y^2 &=& R_{\rm in}^2,  \text{ if } \sqrt{(x-x_0)^2 + y^2}
               >\frac{w}{\sqrt{2}}.  \nonumber
\end{eqnarray}
This curve defines the interface geometry of the inner species in the TF 
approximation. Surface effects, however, smoothen the density profile around 
the meeting point of the two regions as is shown in Fig. \ref{plot2}(b). The 
obstacle beam is repulsive and expels the condensate around it such that bulk 
density is regained over a distance of one healing length.

\subsection{Fully immersed obstacle beam}

 A critical requirement to form coreless vortices is complete immersion 
of the obstacle beam within $n_1$. Based on the previous discussions, as the 
beam approaches the origin, the last point of contact between the beam and 
interface at $R_{\rm in}$ lies along $x$-axis. To determine the condition when 
complete immersion occurs, consider the total potential along $x$-axis
\begin{eqnarray}
 V(x,0,t) &\approx & \frac{x^2}{2} + V_0(t) \left[1 -2\frac{(x-x_0(t))^2}{w^2}
                     \right.  \nonumber \\
           && \left .  + 4\frac{(x-x_0(t))^4} {w^4}\right],
\end{eqnarray}
where, the Gaussian beam potential is considered upto the second order term.
The expression is appropriate in the neighborhood of the beam and has one
local minima and maxima each. There is also a global minima, however, it is 
not the correct solution as it lies in the domain where 
$x>\frac{w}{\sqrt{2}}$. Correct global minima is the one
located at $x=0$ and associated with the harmonic potential. The obstacle is 
considered well immersed when the local minima is located at the interfacial 
radius $R_{\rm in}$. The local minima is a root of the polynomial 
\begin{equation}
x^3+a_2 x^2 +a_1 x+ a_0 = 0,
\end{equation}
obtained from the condition $dV/dx=0$. Here,
\begin{eqnarray}
a_2 &=& -3 x_0, \nonumber\\
a_1 &=& -\frac{w^2}{2}+3 x_0^2+\frac{w^4}{8 V_0}, \nonumber\\
a_0 &=& \frac{1}{2} x_0 \left(w^2-2 x_0^2\right).    \nonumber
\end{eqnarray}
For the trapping potential parameters considered in the present work, all the
three roots are real. The root corresponding to local minima is
\begin{equation}
x_{\rm min}(t)  = -\frac{\left(s_++s_-\right)}{2}-\frac{a_2}{3} 
               + i \frac{\sqrt{3}}{2} \left(s_+-s_-\right),
\end{equation}
where
\begin{eqnarray}
   s_{\pm}&=&\Bigg[-\frac{a_2^3}{27}+\frac{1}{6} \left(-3 a_0+
        a_1 a_2\right)\pm   \nonumber\\
   & &\left.\sqrt{\left(\frac{a_1}{3}-\frac{a_2^2}{9}\right)^3+
       \left(-\frac{a_2^3}{27}+\frac{1}{6} \left(-3 a_0+a_1 a_2\right)\right)
       ^2}\right]^{1/3}. \nonumber
\end{eqnarray}
and $x_{\rm min}$ is a function of time as the obstacle potential is time
dependent. However, for compact notations, like in $V_0$ and $x_0 $, hereafter 
we drop the explicit notation of time dependence while writing 
$x_{\rm min}(t)$. Value of $x_{\rm min}$ is real when it satisfies certain 
conditions \cite{mathworld} and these are met for the experimentally realizable 
parameters. The obstacle is then completely immersed when 
$x_{\rm min}=R_{\rm in}$ and let $t_{\rm im}$ denote the time at which this
condition is satisfied. Once the obstacle beam is well inside the inner 
species, within the beam $n_1$ is zero but $n_2$ is nonzero. It then forms a 
second interface layer, assisted by the obstacle beam, which embeds a bubble 
of the $n_2$ within $n_1$. Recollect, the first interface layer is located at 
$R_{\rm in}$ and it is where $n_2$ encloses $n_1$. The second interface, 
unlike the one at $R_{\rm in}$, is a deformed-ellipse and label it as 
$\Gamma$. It touches $x$-axis at two points and $x_{\rm min}$ is the one 
farther from the origin. From TF-approximation, the density of $n_2$ within 
$\Gamma$ is
\begin{equation} 
  n_{2\Gamma} (x, y, t_{\rm im}) = \frac{\mu_2 - V(x, y, t_{\rm im})}{u_{22}},
\end{equation}
where the subscript $\Gamma$ is to indicate the density is within $\Gamma$. At
$t_{\rm im}$, $n_{2\Gamma}$ is in equilibrium with the bulk of $n_2$, and let
the number of atoms in the be $N_{2\Gamma}$, which can be calculated from 
$n_{2\Gamma}$. 

  Except for deviations arising from mean field energy, the second interface 
$\Gamma$ closely follows an equipotential surface.  From the conditions of
pressure balance at two interfaces
\begin{eqnarray}
 \mu_1-\sqrt{\frac{u_{11}}{u_{22}}}\mu_2 & = &
 \frac{R_{\rm in}^2}{2}\left(1-\sqrt{\frac{u_{11}}{u_{22}}}\right),\nonumber\\
 \mu_1-\sqrt{\frac{u_{11}}{u_{22}}}\mu_2 & = &
 V(x_{\rm min}, 0,t)\left(1-\sqrt{\frac{u_{11}}{u_{22}}}
 \right).
\label{pressure_balance}
\end{eqnarray}
The first equation is valid all along the interface at $R_{\rm in}$. However, 
the second is derived at $(r_{\rm min},0)$, but to a good approximation 
the condition is valid along $\Gamma$ as it is close to an equipotential 
surface. The density of the first species at ($x_{\rm min}$,0) is 
\begin{equation}
n_1(x_{\rm min})  = \frac{1}{\sqrt{u_{11}u_{22}}}\left[
         \frac{R_{\rm out}^2 - 2V(x_{\rm min}, 0,t)}{2} \right].
\end{equation}
Similarly density of second species at at ($x_{\rm min}$,0) is
\begin{equation}
 n_2(x_{\rm min})  =  \frac{1}{u_{22}}\left[\frac{R_{\rm out}^2
              - 2V(x_{\rm min}, 0,t)}{2}\right].
\end{equation}
Inside $\Gamma$, where the obstacle potential is dominant, $n_1$ decays 
exponentially. Along the $x$-axis 
\begin{equation}
  n_1(\delta x_-,0) =  n_1(x_{\rm min},0)\exp\left (-\frac{2}{\Lambda_1}
                       \delta x_-\right ),
\end{equation}
where, $\delta x_- $ is the distance of the point measured from 
$\Gamma$. Similarly the density of the second species just outside 
$\Gamma$ is
\begin{eqnarray}
n_2(\delta x_+,0) = n_2(x_{\rm min},0)\exp\left(-\frac{2}{\Lambda_2}
                    \delta x_+\right ),
\end{eqnarray}
where $\delta x_+$ is distance from $\Gamma$ but away from the obstacle
beam, and $\Lambda_2 $ is the penetration depth of the second species. In 
these expressions the  penetration depth is
\begin{equation}
    \Lambda_i = \xi_i\left [ \frac{\sqrt{a_{11}a_{22}}}
          {a_{12}-\sqrt{a_{11}a_{22}}}\right ] ^{1/2}.
\end{equation}
Based on the densities at the interface region, we can incorporate the 
interspecies interaction and calculate the effective potential of each species. 

\subsection{Obstacle assisted bubble}

 At a time $\Delta t$ after the obstacle is immersed in $n_1$, the location 
and amplitude of the obstacle potential are
\begin{eqnarray}
  x_0(t_{\rm im}+\Delta t) & = & -R_{\rm out} + v_{\rm ob}(t_{\rm im} 
                               + \Delta t),  \\
  V_0(t_{\rm im}+\Delta t) & = & V_0(0) + v_{\rm ob}(t_{\rm im} + \Delta t).
\end{eqnarray} 
Equilibrium TF $n_2$ within the obstacle potential at this instant of time is
\begin{equation} 
  n_{2\Gamma} (x, y, t_{\rm im} + \Delta t ) = \frac{\mu_2 
          - V(x, y, t_{\rm im} + \Delta t)}{u_{22}}.
 \label{n2_den}
\end{equation}
This, however, is higher than the density distribution at $t_{\rm im}$, that is
$n_{2\Gamma} (x, y, t_{\rm im} + \Delta t )> n_{2\Gamma} (x, y, t_{\rm im})$ as
the potential $ V $ is lower. This is on account of two factors: first,
the amplitude of the obstacle potential decreases with time; and second, the
harmonic oscillator potential is lower at $x_0(t_{\rm im}+\Delta t)$. The
number of atoms, however, does not change from the value at $t_{\rm im}$ 
unless there is a strong Josephson current. Density $n_{2\Gamma}$ is thus
below the equilibrium value once the obstacle beam is well within $n_1$. 
This creates a stable bubble of $n_2$ assisted or trapped within the beam
and is transported through the $n_1$.

 Departure of $n_2$ from the equilibrium is not the only density evolution
within the beam. There is a progressive change of $n_{1\Gamma} $ as the beam 
moves deeper into $n_1$. At time $t_{\rm im}$, when the obstacle is 
completely immersed in $n_1$ the effective potential, experienced by $n_1$, 
$V(x, y, t_{\rm im}) + n_{2\Gamma}u_{12}$ is larger than 
$\mu_1$. So, $n_1$ is zero within the beam. However, if the rate of ramping 
$\eta$ is such that at a later time 
$V(x, y, t_{\rm im} + \Delta t) + n_{2\Gamma}u_{12}< \mu_1$, while the beam 
is still within $n_1$, there is a finite $n_1$ within the beam.
\begin{equation}
 n_{1\Gamma} = \frac{\mu_1 - V(x, y, t_{\rm im} + \Delta t)-u_{12}
               n_{2\Gamma}}{u_{11}}
\end{equation}
As the growth of $n_{1\Gamma}$ is in its nascent stage, Eq.~(\ref{n2_den})is 
still applicable. Since $a_{12}>\sqrt{a_{11}a_{22}}$ for the condensate, in 
TF approximation both $n_{1\Gamma}$ and $n_{2\Gamma}$ can not be 
simultaneously non-zero. At the same time, $n_{2\Gamma}$ is forbidden to
migrate to the bulk $n_2$ due to the $n_1$ generated potential barrier in the 
region between interfaces $\Gamma$ and $R_{\rm in}$. To accommodate both
$n_1$ and $n_2$ within the beam, the shape of interface $\Gamma$ is 
transformed to increase $n_{2\Gamma}$. So that $n_2$ is zero in certain regions 
within the beam where the condition 
$V(x, y, t_{\rm im} + \Delta t) + n_{2\Gamma}u_{12}<\mu_1$ is satisfied. This 
mechanism is responsible for obstacle assisted transport of $n_2$ across $n_1$.

\subsection{Obstacle induced density instability }

 The harmonic potential increases once the obstacle crosses origin, but the 
obstacle potential continues to decrease. A fine balance is achieved if
\begin{equation}
  \frac{dV_{\rm obs}}{dx} = \frac{\partial V_{\rm obs}}{\partial x} + 
     \left ( \frac{dx}{dt}\right )^{-1}\frac{\partial V_{\rm obs}}{\partial t},
\end{equation}
the dynamical rate of change of obstacle potential, is same in magnitude and 
opposite to the rate of change of the harmonic potential $dV_{\rm ho}/dx$. 
It is, however, difficult to satisfy this condition across the entire obstacle 
potential as the Gaussian beam has has different profile than the oscillator 
trapping potential. But, the parameters may be tuned so that the condition is 
satisfied at the center of the beam, then
\begin{equation}
  \frac{dV_{\rm ho}}{dx}  =  \frac{\eta}{v_{\rm obs}}.
\label{temp}
\end{equation}
In the present case $\eta$ is a constant and the 
condition in Eq. (\ref{temp}) is satisfied only when the beam is at a 
specific point $x_{\rm e}$, then
\begin{equation}
\left . \frac{dV_{\rm ho}}{dx}\right|_{x_{\rm e}}  =  \frac{\eta}{v_{\rm obs}}
\label{x_e}
\end{equation}
Before reaching $x_{\rm e}$, the potential within the beam decreases, but 
increases after crossing it. At some point beyond $x_{\rm e}$, the total
potential experienced by $n_2$ bubble can become greater than $\mu_2$, 
resulting in an unstable bubble.


\section{Stability of coreless vortex dipole}

  To analyze the energetic stability of coreless vortex dipoles in
phase-separated binary condensates, we adopt the following {\em ansatz} for
the wave function of inner and outer species 
\begin{eqnarray}
  \psi_1(x,y) & = & a\left[i \left(x - v_1\right) + \left(y^2 
                  - v_2^2\right)\right]\exp \left( - \frac{x^2 
                  + y^2}{2}\right),\nonumber\\
\psi _2(x,y) & = & b\left(x^2 + y^2\right)\exp\left( - \frac{x^2 
                  + y^2}{2}\right),
\label{ansatz}
\end{eqnarray}
where $(v_1,\pm v_2)$ is the location of the vortex dipole. With this choice
of {\em ansatz}, the vortex dipole in the inner species is a normal vortex 
dipole. For simplicity of analysis, consider the trapping potential is
purely harmonic, i.e.
\begin{equation}
  V(x,y)  = \frac{x^2+y^2}{2}.
\end{equation}
Normalization yields the following constraints on $a$ and $b$
\begin{eqnarray}
  a & = & \left [ \frac{4}{5\pi+4\pi (v_1^2- v_2^2+ v_2^4)} \right ] ^{1/2},
                  \nonumber\\
  b & = &\sqrt{ \frac{1}{2\pi} },
\label{b}
\end{eqnarray} 
and reduces the number of unknowns to two $v_1$ and $v_2$. The energy of the 
binary condensate using the {\em ansatz} in Eq.~({\ref{ansatz}}) is
\begin{eqnarray}
E & = &\frac{N_2 b^2}{\pi} \left(4 +\frac{3}{8} b^2 u_{22}\right)+
      \frac {N_1a^2\pi}{1024} \left\lbrace 32 \left[8 (11\right.\right.
      \nonumber\\
  &   & \left.+4 v_1^2-4 v_2^2+4 v_2^4)+b^2 u_{12} \left(15+8 v_1^2-12 v_2^2
        +8 v_2^4\right)\right]\nonumber\\
  &   &  +a^2 u_{11} \left[177-304 v_2^2+256 v_1^4+416 v_2^4-256 v_2^6
        \right.\nonumber\\
  &   & \left.\left.+256 v_2^8+32 v_1^2 \left(15-8 v_2^2+16 v_2^4\right)
        \right]\right\rbrace.
\label{energy}
\end{eqnarray}
In the same way, we can also determine the chemical potentials of the two 
species in terms of the same parameters. 
The stationary points of the energy function are the solutions of the coupled
equations $ \partial E/\partial v_1 =  0$ and $\partial E/ \partial v_2= 0$.
These equations are solvable analytically, but the solutions are lengthy and
cumbersome. The solution of particular interest is the vortex dipole located
along the $y$-axis ($v_1=0$). This is consistent with the observation that,
regardless of the orientation, from the symmetry the axis passing through 
vortex dipole can be considered as the $y$-axis. A rotation
transformation is sufficient to achieve this. For the case considered, the 
energy extrema conditions simplify to equations

\begin{eqnarray}
\frac{\partial E}{\partial v_1} 
     & = & \frac{\pi a N_1}{256} \bigg\{16 \left[8   
           \left(11-4 v_2^2+4 v_2^4\right)+b^2 u_{12} \left(15-12 v_2^2
           \right.\right.\nonumber\\
     &   & \left.\left.+8 v_2^4\right)\right] \frac{\partial a}{\partial v_1}
           +  a{}^2 u_{11} \left[177-304 v_2^2+416 v_2^4\right.\nonumber\\
     &   & \left.-256 v_2^6+256 v_2^8\right] 
           \frac{\partial a}{\partial v_1}\bigg\},\nonumber\\
\frac{\partial E}{\partial v_2} & = &\frac{\pi a N_1}{256}  
           \bigg\{64 a v_2 \left[8 \left(2 v_2^2 - 1\right)+b^2 u_{12} 
           \left(4 v_2^2 - 3\right)\right]\nonumber\\
     &   & -8   a {}^3 u_{11} v_2 \left[19-52 v_2^2+48 v_2^4-64 v_2^6
            \left(4 v_2^2-1\right)\right]\nonumber\\
     &   & +16 \left[8 \left(11
           -4 v_2^2+4 v_2^4\right)+b^2 u_{12} \left(15-12 v_2^2
            \right.\right.\nonumber\\
     &   & \left.\left.+8 v_2^4\right)\right]  
            \frac{\partial a}{\partial v_2}+a{}^2 u_{11}\left[177-304 v_2^2
            +416 v_2^4\right. \nonumber\\
     &   & \left.-256 v_2^6+256 v_2^8\right] 
           \frac{\partial a}{\partial v_2}\bigg\}.
\end{eqnarray}
For the sake of illustration, consider, $N_1 = 10^5$, $N_2 = N_1/2$,
$a_{11} = 0.99a_0$, $a_{22} = 4.6a_0$, $a_{12} = 2.14a_0$, $\beta = 11.25$,
$\omega/2\pi=8$, and $V_0(t) = 0$. Call this set of parameters as parameters 
{\em set b}. The stationary points corresponding to minimum energy for this 
parameters set are $(v_1,v_2) = (0,\pm 0.9678)$.

\subsection{Coreless vortex dipole}

To compare the energy of this normal vortex dipole with that coreless 
vortex dipole, consider that the cores of the vortices at $(v_1,v_2)$ and 
$(v_1,-v_2)$ are filled with outer species. We approximate the cores as 
circular with a radius of $\xi = 1/\sqrt{2\mu_1}$, the 
coherence length of the inner species. We approximate the density profiles 
of the outer species within the cores with the TF distribution
\begin{equation}
 \psi_2(x,y) = \left \{ 
      \begin{aligned}
         & \sqrt{\frac{\mu_2 - V(x,y)}{u_{22}}} & r_c^2 \leqslant\xi^2 \\
         & b\left(x^2+y^2\right)\exp\left(-\frac{x^2+y^2}{2}\right) 
                 &  r_c^2>\xi^2,               
      \end{aligned} \right .
\end{equation}
where $r_c = \sqrt{(x-v_1)^2+(y \pm v_2)^2}$ is the radial distance from the 
core of each vortex and antivortex. The above expression for $\psi_2$ is
renormalized to obtain the modified value of $b$, which in turn enables us
to calculate the energy for the coreless vortex dipole. To obtain the 
analytic expression for energy, we integrate $b\left(x^2+y^2\right)\exp
\left(-\frac{x^2+y^2}{2}\right)$ over the whole real space without excluding 
the core regions. The modified value of $b$ after the normalization is
\begin{equation}
b \approx \left [ \frac{2u_{22}- \pi  \xi ^2 \left(\xi ^2+2 v_1^2+
               2 v_2^2-4 \mu _2\right)}{4\pi u_{22}}\right ]^{1/2}.
\label{modified_b}
\end{equation} 
With this approximation, the number of atoms of outer species embedded within
the cores is
\begin{equation}
   \delta N_2 = N_2 \frac{\pi  \xi ^2 \left(\xi ^2 + 2 v_1^2 + 2 v_2^2 
                - 4 \mu _2\right)} {2 u_{22}}.
\end{equation}
These contribute to the total energy through the potential and intra-species 
interaction energies. The energy contribution from the cores is  then
\begin{eqnarray}
\delta E & = & -\frac{\pi  \xi ^2}{12 u_{22}} \left[\xi ^4+3 v_1^4+6 \xi ^2 
               v_2^2+3 v_2^4+6 v_1^2\left(\xi ^2+v_2^2\right)\right.
               \nonumber\\
         &    &\left.-12 \mu _2^2\right].
\end{eqnarray}
The total energy of the binary condensate with coreless vortex is
\begin{equation}
 E_c = E + \delta E,
\end{equation} 
where $E$ is defined by Eqs.~(\ref{energy}) and (\ref{modified_b}).
For the parameter {\em set b}, the energy per particle [$E_c/(N_1+N_2)$] is 
$4.18$ which is smaller than the value of $4.26$ obtained from 
Eqs.~(\ref{b}) and (\ref{energy}). This implies
that coreless vortex dipole is energetically more favorable than normal 
vortex dipole with the percentage energy difference of $1.9\%$. 

 The Gaussian form of the {\em ansatz} imposes limitations on the domain of
applicability. But, the ansatz provide qualitative information on the 
stability coreless vortex dipoles. To relate to experimental realizations, 
consider $^{85}$Rb-$^{87}$Rb binary condensate with $a_{11} = 460a_0$,  
$a_{22} = 99a_0$, and $a_{12} = 214a_0$ as the scattering length values 
and $2 N_1 = N_2 = 10^6$ as the number of atoms. Here $a_{11}$ is tunable 
with magnetic Feshbach resonance \cite{Cornish}. With this set of parameters, 
the stationary state of $^{85}$Rb-$^{87}$Rb binary condensate is just 
phase-separated. The trapping potential and obstacle laser potential 
parameters are same as those considered in 
Ref.\cite{Neely}, that is $\omega/(2\pi) = 8$Hz, $\alpha = 1$, $\beta = 11.25$,
$V_0(0) = 93.0\hbar\omega$, and $w_0 = 10\mu$m. Hereafter we term this set of 
scattering lengths, number of atoms, and trapping potentials as 
{\em set a}.  For this parameter set, in absence of
the obstacle potential, the stationary point of minimum energy is
$(v_1,v_2) = (0,\pm 0.75275)$. And, the difference in the energies per
particle of the binary condensate with normal and coreless vortex dipoles is 
$0.11$.


\subsection{TF approximation single vortex}

 As mentioned earlier, the applicability of the variational {\em antsatz} is 
limited to the weakly interacting domain. Experiments, however, mostly explore
domains other than weakly interacting. To analyze stability in this domain, we 
resort to the TF approximation. 

\subsubsection{Normal vortex}

 To simplify the analysis, consider a normal vortex located at the 
center of the inner species. For a normal vortex, the wave functions of
the first or inner species  in TF approximation is
\begin{equation}
 \psi_1(r)  =  \left \{
     \begin{aligned}
        & \sqrt{\frac{\mu _1-V(r)}{u_{11}}}\frac{r}{\sqrt{r^2+\xi^2}}
                 &r^2 \leqslant R_{\rm in}^2 , \\
        & 0      &r^2 > R_{\rm in}^2 . 
      \end{aligned} \right .
\end{equation}
Similarly, the wave function of the second  or outer species is
\begin{equation}
 \psi_2(r)  =  \left \{
      \begin{aligned}
         & \sqrt{\frac{\mu _2-V(r)}{u_{22}}}
                 &R_{\rm in}^2\leq r^2\leq R_{\rm out}^2\\
         & 0     &r^2 > R_{\rm out}^2
      \end{aligned} \right . 
\end{equation}
here $V(r)$, $R_{\rm in}$, $R_{\rm out}$, and $\xi$ are the trapping potential,
radius of the
interface between inner and outer species, radial extent of outer species, and
coherence length of inner species, respectively. The chemical potentials
$\mu_1$ and $\mu_2$ are parameters of the calculations and obtained from 
inverting the normalization condition
\begin{eqnarray}
   N_1 & = & \frac{\pi}{8 u_{11} \mu _1^2}\left[2 R_{\rm in}^2 \left(\mu _1
             +4 \mu _1^3\right)- 2 R_{\rm in}^4 \mu _1^2+\ln \left(1
             +2 R_{\rm in}^2 \mu _1\right) \right.
                            \nonumber   \\
         & & \left. \times \left(1+4 \mu _1^2\right) \right]\label{N_1&N_2_1}
                                        \\
   N_2  & = & \frac{\pi  \left(R_{\rm in}^2-2 \mu _2\right){}^2}{4 u_{22}}.
\label{N_1&N_2_2}
\end{eqnarray} 
The total energy of the vortex with a normal vortex located at origin is
\begin{eqnarray}
E & = &\int \left\lbrace \psi_1(r)\left[-\frac{1}{r}\frac{\partial}{\partial r}-
        \frac{\partial^2}{\partial r^2}+\frac{1}{r^2}+V(r) + \frac{u_{11}}{2}
        |\psi_1(r)|^2\right]\right.\nonumber\\
  &   & \left.\psi_1(r) + \psi_2(r) \left[V(r) + \frac{u_{22}}{2}|\psi_2(r)|^2 \right]
       \psi_2(r)\right\rbrace,
\end{eqnarray}
The energy can be minimized under the constraint that $\mu_1$ and $\mu_2$ 
satisfy Eqs.~(\ref{N_1&N_2_1}-\ref{N_1&N_2_2}) to completely determine the 
stationary state of binary condensate with a normal vortex at the origin.


\subsubsection{Coreless Vortex}
For a coreless vortex, the wave functions of the two species in TF 
approximation are 
\begin{eqnarray}
 \psi_1(r)  &=& \left \{
     \begin{aligned}
        & \sqrt{\frac{\mu _1-V(r)}{u_{11}}}\frac{r}{\sqrt{r^2+\xi^2}}
                 &r^2 \leqslant R_{\rm in}^2\\
        & 0&r^2 > R_{\rm in}^2,
     \end{aligned} \right . \\
 \psi_2(r)  &=&  \left \{
     \begin{aligned}
        & \sqrt{\frac{\mu _2-V(r)-u_{12}\psi_1(r)^2}{u_{22}}} 
                      && r^2 \leqslant \xi^2\\
        &  \sqrt{\frac{\mu _2-V(r)}{u_{22}}} 
                      &&R_{\rm in}^2\leqslant r^2\leqslant R_{\rm out}^2\\
        &  0          &&r^2 > R_{\rm out}^2.
     \end{aligned} \right .
\end{eqnarray}
Normalizing the wave functions gives the constraint equations in terms
of $\mu_1$ and $\mu_2$ as 
\begin{eqnarray}
N_1 & = & \frac{\pi}{8 u_{11} \mu _1^2}\left[2 R_{\rm in}^2 \left(\mu _1+4 
          \mu _1^3\right)- 2 R_{\rm in}^4 \mu _1^2+\ln \left(1+2 R_{\rm in}^2 
          \mu _1\right) \right.\nonumber\\
    &   & \left. \times\left(1+4 \mu _1^2\right) \right],\nonumber\\
N_2 & = & \frac{\pi}{16 u_{11} u_{22} \mu _1^2}  \left\{u_{21} \left[-1
          +\ln 4+8 (\ln 2-1) \mu _1^2\right]+u_{11}\right.\nonumber\\ 
    &   & \left.\times\left(4 R_{\rm in}^4 \mu _1^2+8 \mu _1 \mu _2
           -16 R_{\rm in}^2 \mu _1^2 \mu _2+16 \mu _1^2 \mu _2^2-1\right)
          \right\}.
\end{eqnarray}
From the expressions, we can calculate the total energy of the system. It is 
a complicated expression which we refrain from writing. For the 
present case, it is perhaps sufficient to say, it is possible to obtain an
analytical but cumbersome expression of total energy.

\begin{figure}[ht]
\includegraphics[width=8.5cm] {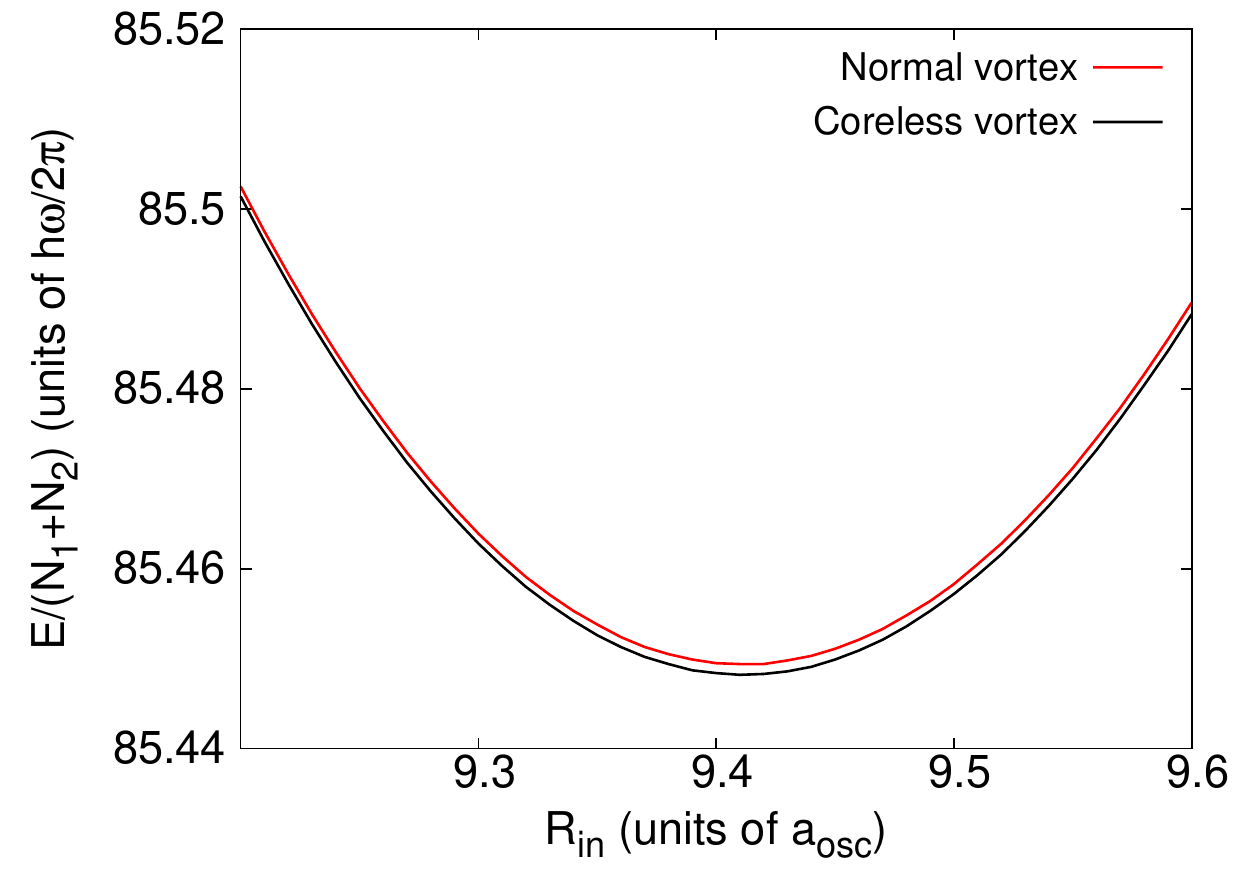}
\caption{The energy of the binary condensate with $V_0=0$ and rest of the
parameters same as those in the parameters {\em set a} as a function of
$R_{\rm in}$. The condensate has a vortex dipole located at origin.
Black and blue curves are for coreless and normal
vortex dipoles respectively.}
\label{fig_coreless_vortex}
\end{figure}

\begin{figure}[ht]
\includegraphics[width=8.5cm] {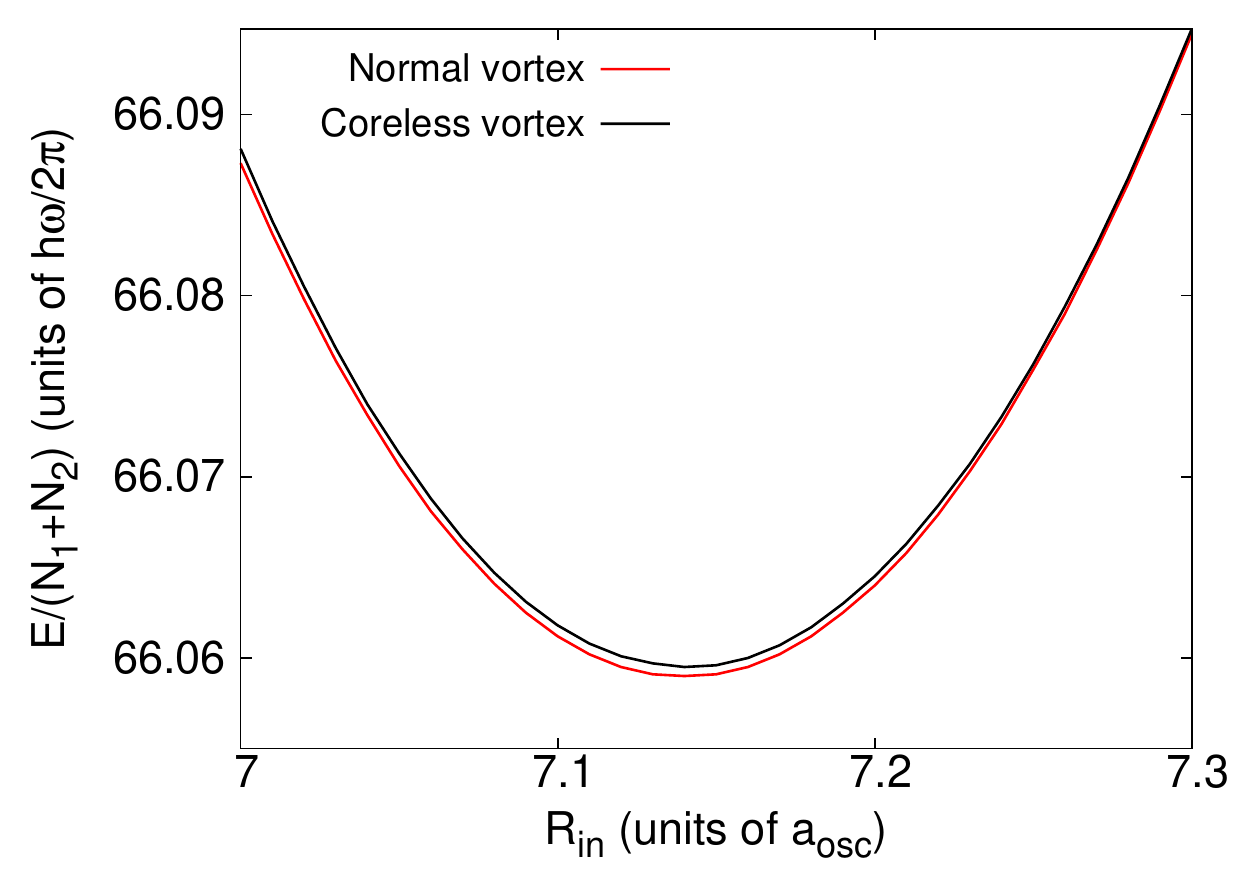}
\caption{The energy of the binary condensate with $N_1=N_2=10^6$,
$a_{11}=51a_0$, $a_{22}=99a_0$, $V_0 = 0$, and rest of the parameters same as
those in parameters {\em set a} as a function of $R_{\rm in}$. The condensate
has a vortex located at the origin. Black and blue curves are for coreless and
normal vortex dipoles respectively.}
\label{fig_normal_vortex}
\end{figure}

For the parameter {\em set a} with $V_0(t) = 0$, binary 
condensate with the coreless vortex has lower energy than the one with normal 
vortex. This is evident from Fig.\ref{fig_coreless_vortex} where the variation 
of total energy is shown as a function of $R_{\rm in}$. 
However, the stability 
vanishes if one considers $N_1=N_2=10^6$, $a_{11}=51a_0$, $a_{22}=99a_0$ in 
{\em set a}. It is then the normal vortex which is energetically stable. This 
is evident form Fig.\ref{fig_normal_vortex}.
 These results are in very good 
agreement with numerical results. This TF analysis, albeit, is for single
vortex clearly illustrates that depending upon the interaction parameters, a 
phase-separated binary condensate can either support coreless or normal 
vortex. It must be emphasized, the energy, chemical potentials, and interface 
radius obtained from the TF approximation are in very good agreement 
with numerical solution of coupled GP equations. For example, the numerical 
value of energy $66.32$ corresponding to Fig.\ref{fig_normal_vortex} is very 
close to the value $66.06$ from TF approximation.


\subsection{TF approximation with vortex dipole}

 We use the TF approximation to examine the stability of coreless vortices 
in the strongly interacting domain. 

\subsubsection{Normal vortex dipole}
Assume the vortex affects the density of the condensate only  within the core 
regions. We then adopt the following {\em ansatz} for binary condensate with a 
normal vortex dipole 
at $(v_1,\pm v_2)$.
\begin{eqnarray}
 \psi_1(r)  & = & \left \{ 
     \begin{aligned}
        & 0 \!&& x^2+y^2 > R_{\rm in}^2\\
        & 0 \!&& [(x-v_1)^2+(y\pm v_2)^2] \leqslant \xi^2\\
        & \sqrt{\frac{\mu_1 - V(x,y)}{u_{11}}} 
              && \left \{ 
                   \begin{aligned}
                     & x^2+y^2 \leqslant R_{\rm in}^2~\&\\
                     &~[(x-v_1)^2+(y\pm v_2)^2] > \xi^2
                  \end{aligned} \right .
     \end{aligned} \right .   \\
 \psi_2(r)  &=&  \left \{ 
     \begin{aligned}
        & \sqrt{\frac{\mu _2-V(x,y)}{u_{22}}}
                 &&R_{\rm in}^2\leqslant (x^2+y^2) \leqslant R_{\rm out}^2\\
        & 0&&(x^2+y^2) > R_{\rm out}^2\\
        & 0&&(x^2+y^2) < R_{\rm in}^2.
     \end{aligned} \right .
\end{eqnarray}
The vortex dipole contributes mainly through the kinetic energy of $\psi_1$, 
which may be approximated with the  value of single species condensate given 
in Ref.\cite{Zhou}
\begin{equation}
   E_{\rm vd} = \frac{2\mu _1}{u_{11}}\ln\left(\frac{2v_2}{\xi}\right).
\end{equation}
Using these {\em ansatz} the number of atoms are
\begin{eqnarray}
N_1 & = & \frac{\pi  \left(1+4 v_1^2 \mu _1+4 v_2^2 \mu _1-8 
          \mu _1^2-2 R_{\rm in}^4 \mu _1^2+8 R_{\rm in}^2 
          \mu _1^3\right)}{8 u_{11} \mu _1^2},\nonumber \\
N_2 & = & \frac{\pi  \left(R_{\rm in}^2-2 \mu _2\right){}^2}{4 u_{22}}.
\end{eqnarray}
In a similar way, we can evaluate the energy of the entire condensate.


\subsubsection{Coreless vortex dipole}
\begin{center}
\begin{figure}[ht]
\includegraphics[width=8.5cm] {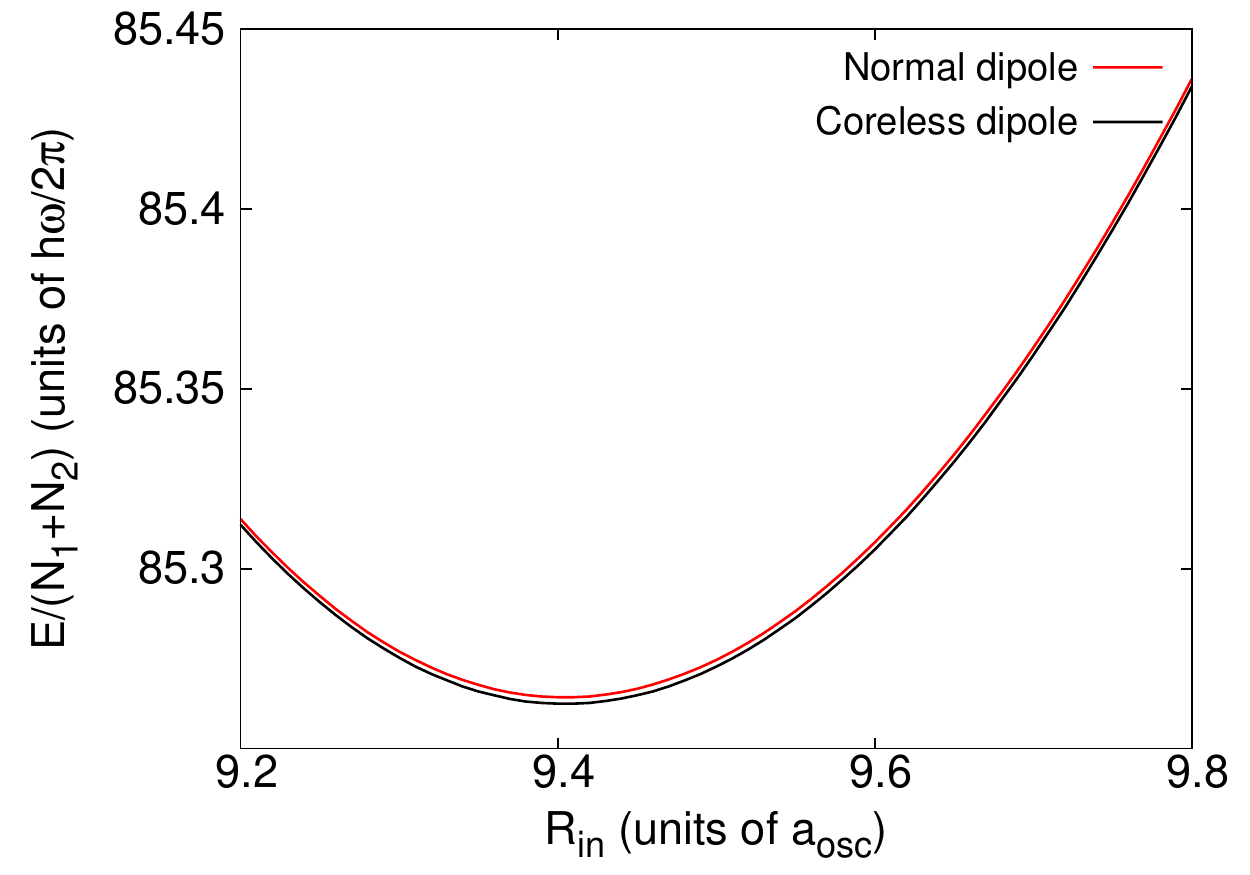}
\caption{The energy of the binary condensate with $V_0=0$ and rest of the
parameters same as those in the parameters {\em set a} as a function of
$R_{\rm in}$. The condensate has a vortex dipole located at $(0,\pm 1)$.
Black and blue curves are for coreless and normal vortex dipoles
respectively.}
\label{fig_coreless_vortex_dipole1}
\end{figure}
\end{center}

\begin{figure}[ht]
\includegraphics[width=8.5cm] {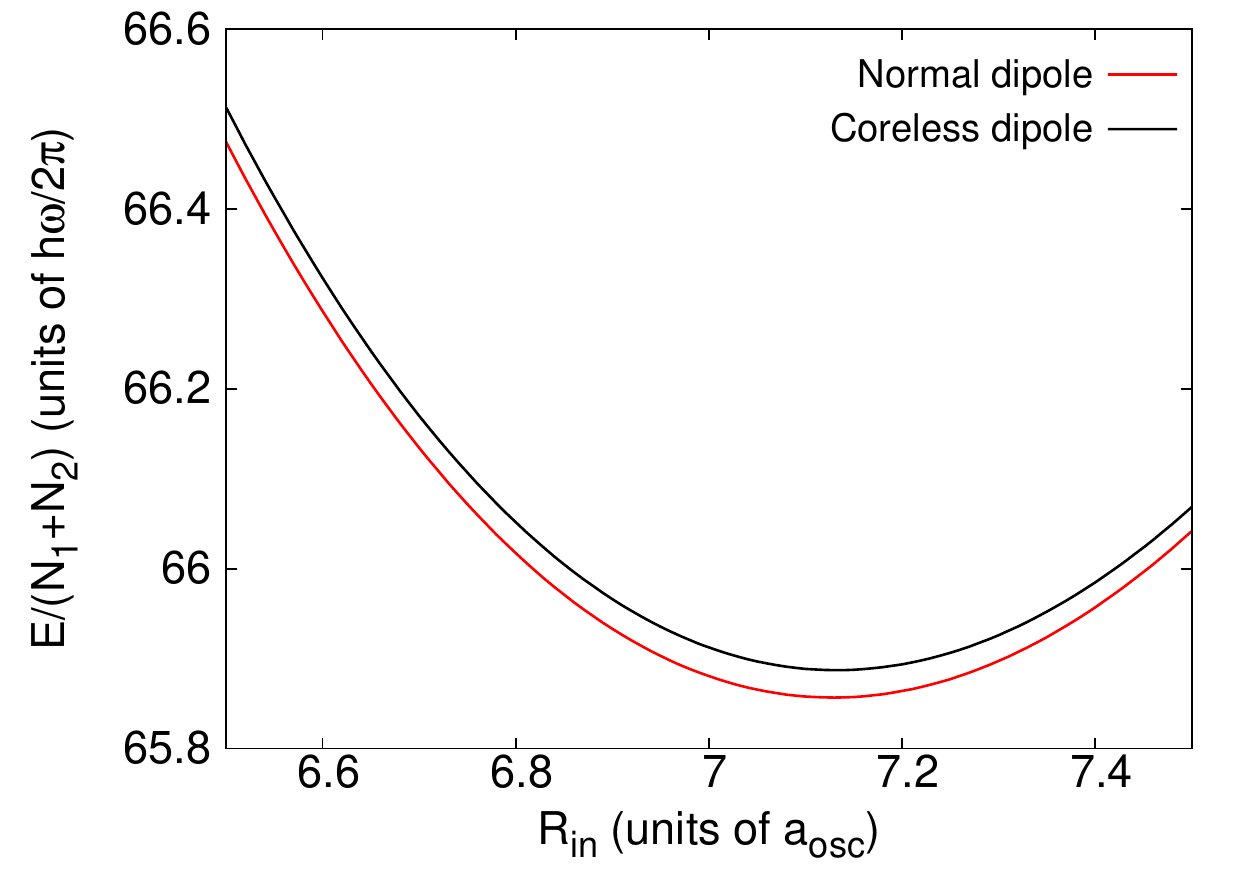}
\caption{The energy of the binary condensate with $N_1=N_2=10^6$, 
$a_{11}=51a_0$, $a_{22}=99a_0$, $V_0 = 0$, and rest of the parameters same as 
those in parameters {\em set a} as a function of $R_{\rm in}$. The condensate 
has vortex dipole located at $(0,\pm 1)$. Black and blue curves are for 
coreless and normal vortex dipoles respectively.}
\label{fig_coreless_vortex_dipole2}
\end{figure}

For coreless vortex dipole, we adopt the {\em ansatz}
\begin{eqnarray}
 \psi_1(r)  & = &  \left \{
      \begin{aligned}
         & 0 && x^2+y^2 > R_{\rm in}^2\\
         & 0 && [(x-v_1)^2+(y\pm v_2)^2] \leqslant \xi^2\\
         & \sqrt{\frac{\mu_1 - V(x,y)}{u_{11}}} 
              && \left \{ 
                \begin{aligned}
                   & x^2+y^2 \leqslant R_{\rm in}^2~\&\\
                   &~[(x-v_1)^2+(y\pm v_2)^2] > \xi^2,
                \end{aligned} \right .
      \end{aligned} \right . \\
 \psi_2(r)  & = & \left \{
      \begin{aligned}
         & \sqrt{\frac{\mu _2-V(x,y)}{u_{22}}}
                &&\left \{
                   \begin{aligned}
                      & R_{\rm in}^2\leqslant (x^2+y^2) \leqslant R_{\rm out}^2~||~\\
                      & [(x-v_1)^2+(y\pm v_2)^2] \leqslant \xi^2
                   \end{aligned} \right . \\
         & 0&&(x^2+y^2) > R_{\rm out}^2\\
         & 0&& \left \{
                \begin{aligned}
                   & x^2+y^2 < R_{\rm in}^2~\&\\
                   &~[(x-v_1)^2+(y\pm v_2)^2] > \xi^2.
                \end{aligned} \right .
      \end{aligned} \right .
\end{eqnarray}
Using these {\em ansatz} the modified expressions for $N_2$ is
\begin{eqnarray}
N_2 &=& \frac{\pi}{8 u_{22} \mu _1^2}\left(2 R_{\rm in}^4 \mu _1^2
       + 8 \mu _1 \mu _2-8 R_{\rm in}^2 \mu _1^2 \mu _2+8 \mu _1^2 \mu _2^2 -1 
         \right.       \nonumber \\
    & &\left . -4 v_1^2 \mu _1-4 v_2^2 \mu _1\right).
\end{eqnarray}
We can also calculate the total energy $ E$ of the system. The important 
change in $E$ is the inclusion of interface interaction energy $E_{\rm int}$.
It arises from the interface interactions at the cores of the vortex and 
antivortex. Based on Ref.~\cite{Timmermans},
\begin{equation}
 E_{\rm int} = \frac{8}{3}Pb\pi\xi\left (\frac{a_{12}}{\sqrt{a_{11}a_{22}}}-1
               \right ),
\end{equation}
where $P$ is the pressure on the circumference of the cores and 
\begin{equation}
   b = 2\left [ \frac{3(\mu_1+\mu_2)\sqrt{a_{11}a_{22}}}{4\mu_1\mu_2
          (a_{12}-\sqrt{a_{11}a_{22}})} \right ] ^{1/2}.
\end{equation}
As in the previous section, energy can be minimized with the constraint
of the fixed number of atoms. For the parameters {\em set a}  
without obstacle potential, the coreless vortex dipole has lower energy than 
the normal vortex dipole and is shown in 
Fig.~\ref{fig_coreless_vortex_dipole1} for the vortex dipole located
at $(0,\pm1)$.
 For $N_1=N_2=10^6$, $a_{11}=51a_0$, $a_{22} = 99a_0$ and rest
of the parameters same as in parameter {\em set a}, condensate with the normal
vortex dipole has lower energy than the one with coreless vortex dipole
(see Fig.~\ref{fig_coreless_vortex_dipole2}). These results are again in good 
agreement with the numerical results.


\section{Numerical results}

   For the numerical calculations, we again resort to experimentally realizable
system and paramteres. We again work with the $^{85}$Rb-$^{87}$Rb mixture and
parameter {\em set a} introduced earlier. Recall, the  values of paramteres in
this set are $a_{11} = 460a_0$, $a_{22} = 99a_0$, and $a_{12} = 214a_0$ 
$2 N_1 = N_2 = 10^6$, $\omega/(2\pi) = 8$Hz, $\alpha = 1$, $\beta = 11.25$, $
V_0(0) = 93.0\hbar\omega$, and $w_0 = 10\mu$m.
As mentioned earlier, $a_{11}$ is tunable with magnetic Feshback resonance 
\cite{Cornish}. 

   The transport of $n_2 $ within the region of influence of $V_{\rm obs}$ 
is crucial to generate coreless vortices. As 
$V_{\rm obs}$ enters $n_1$, it nudges $n_1$ and creates an interstice, which
is occupied by $n_2$. This is an outcome of the difference in 
coherence lengths ($\xi_i = 1/\sqrt{2n_iu_{ii}}$). The smaller $\xi$ of $n_2$
ensures that it recovers bulk value over shorter distances. It implies that it 
is comparatively more difficult to displace the $n_2$ than $n_1$. As the 
obstacle reaches the interface with diminished strength, it is unable to 
displace $n_2$, but not for $n_1$ due larger $\xi$. This results in an inward 
protrusion of the interface, which detaches from the interface at a later time. 
For detachment to occur, it should be energetically favorable. The 
difference in $\xi$ coupled with $a_{12}\gtrapprox\sqrt{a_1a_2}$ ensures that 
the interstitial filling of $n_2$ is energetically lower, in other words 
energetically favorable. This is easily confirmed with imaginary time 
propagation of GP equation to obtain the stationary state geometry of the 
binary condensate with $V_{\rm obs} $ located inside $n_1 $ as is shown in 
Fig.~\ref{plot2}(c). 
\begin{figure}
  \includegraphics[width=8.5cm] {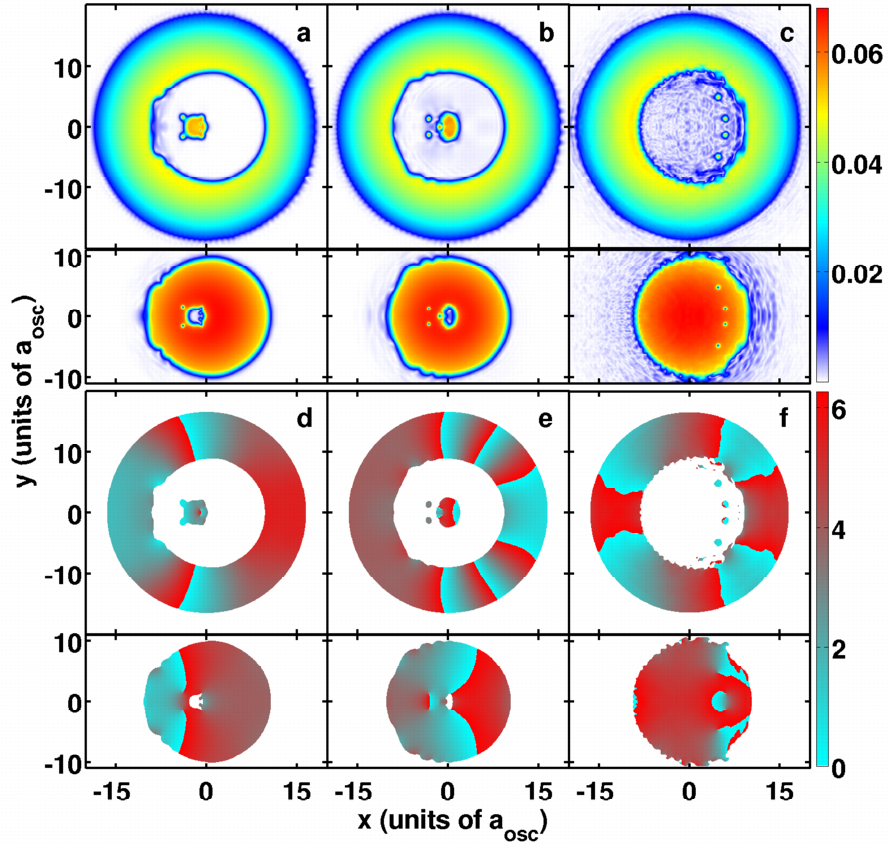}
  \caption{Coreless vortex dipole shedding when the obstacle potential 
           traverses from the edge to the inner component in a binary 
           condensate with $V_0 = 125$, $x_0(0) = -15a_{\rm osc}$ and rest of 
           the parameters same as those in {\em set a}. Figures a, b and c
           are the absolute values of wave functions at $0.21$s, $0.33$s, 
           $0.56$s, respectively; while figures d, e and f
           are the corresponding phase plots. 
          }
  \label{coreless_vdip}
\end{figure}


\subsection{Motion of $V_{\rm obs} $ towards interface}

   We first consider the motion of $V_{\rm obs} $, initially located in the 
$n_2$ towards the interface. For this case, we consider $^{85}$Rb-$^{87}$Rb 
binary condensate with parameter {\em set a} and $V_{\rm obs}$ of 
$V_0(0) = 125.0$. The $V_{\rm obs} $ is initially located at 
$x = -15 a_{\rm osc}$ and moves with the speed of $180\mu$m/s. It progressively
decreases in strength and vanishes at $x = 8a_{\rm osc}$. The obstacle creates 
a normal vortex dipole as it traverses $n_2$. And, as $V_{\rm obs} $ approaches
the interface, it transports the vortex dipole. Further motion of $V_{\rm obs}$
in the $n_1$  generates coreless vortices in the inner component. The formation
of the coreless vortex dipole is shown in Fig. \ref{coreless_vdip}. 
The plots in the figure show both the density and phase 
distribution. From the phase plot, it is clear that the phase singularity of 
the vortex is associated with $n_1$. And, from the density plot it is evident 
that the vortex-antivortex cores are filled with $n_2$. One effect which 
precedes the seeding of coreless vortex is the indentation of the inner 
interface after it cleaves when the $V_{\rm obs} $ is fully 
immersed in $n_1$. This may be explained in terms of the dynamical evolution 
around $V_{\rm obs} $ and analyzed in two different ways.

\subsubsection{Inertia of $n_2$ }

 Just after the interface cleaves, when the $V_{\rm obs} $ is set 
into motion at $t = t_0$, the $n_2$ due to inertia, does not follow 
$V_{\rm obs}$ initially. As a result at $t_0+\delta t$, neglecting the back 
flow of $n_2$, outer species (along $x$-axis) at $x' \in [x_{\rm min}(t_0), 
x_{\rm min}(t_0+\delta t)]$ is approximately same
\begin{equation}
 n_2(x',0,t+\delta t) \approx n_2(x',0,t_0),
\end{equation}
here $\delta t$ is time taken by the outer species to redistribute itself 
according to the displaced obstacle potential. As a result over a length 
$ x_{\rm min}(t_0+\delta t) - x_{\rm min}(t_0)$, the density of inner species in 
TF approximation is 
\begin{equation}
 n_1(x',0,t_0+\delta t) =  \frac{\mu_1 - x^2/2 -n_2(x',t_0)u_{12}}{u_{11}}.
\end{equation}
Hence inner $n_1$ flows into the above domain and replaces $n_2$. The circular 
symmetry of $V_{\rm obs}$ coupled with motion
along $x$-axis ensures that the aforementioned process starts along $x$-axis
earlier; also the replacement of $n_2$ by $n_1$ takes place at the much faster
speed as compared to the speed of the obstacle. This leads to creation of dimple
on the inner interface.

\subsubsection{Pressure balance}
 Let $x_0$ and $\Gamma_{x_0}$ be the location of the obstacle
just after the interface cleaves and approximately circular inner interface 
centered around $x_0$ with radius $\tilde R_{\rm in}$, respectively. After 
time $\delta t$, let $x_0'= x_0 + \delta x_0$ and $\Gamma_{x_0'}$ be the 
location of the obstacle and inner interface, respectively. Assuming that 
$\Gamma_{x_0'}$ has same radius as $\Gamma_{x_0}$, i.e. $\tilde{R}_{\rm in}$, 
the points of intersection of $\Gamma_{x_0}$ and $\Gamma_{x_0'}$ are
\begin{equation}
(x_c,y_c) = \left(\frac{1}{2} (x_0+ x_0'),
            \pm \frac{1}{2} \sqrt{4 {\tilde{R}_{\rm in}}^2-
            \delta x_0^2}\right).
\label{cross_over}
\end{equation}
Thus the region $\delta \Gamma$ defined as
\begin{eqnarray}
 (x-x_0)^2+y^2  <  \tilde{R}_{\rm in}^2<(x-x_0')^2+y^2
\label{exp_region}
\end{eqnarray}
with area 
\begin{equation}
A = \pi \tilde{R}_{\rm in}^2 - 2\tilde{R}_{\rm in}^2\cos^{-1}
     \frac{\delta x_0}{2\tilde{R}_{\rm in}}+ \frac{\delta x_0}{2}
     \sqrt{4\tilde{R}_{\rm in}^2-\delta x_0^2},
\end{equation}
lies within $\Gamma_{x_0}$ but outside $\Gamma_{x_0'}$.
Let $P_1 = n_1^2u_{11}/2$ and $P_2 = n_2^2u_{22}/2$ be the pressures exerted 
by inner and outer species on the left arc of $\delta \Gamma$ region, 
respectively. The potential experienced by $n_1$ along the left arc of $\delta 
\Gamma$, which is approximately equal to harmonic trapping potential, does not 
change as the obstacle is moving away from this region, and so does the $P_1$. 
The potential experienced by $n_2$ over $\delta \Gamma$, which is equal to sum 
total of harmonic trapping and obstacle potentials, decreases as the obstacle 
moves from $x_0$  to $x_0+\delta x_0$. In fact, as one moves from $(x',0)$, 
where $x'\in(x_0-\tilde{R}_{\rm in},x_0+\delta x_0-\tilde{R}_{\rm in})$, to 
the $(x_c,y_c)$ along a curve with same sign of curvature as bounding arcs 
of $\delta \Gamma$ region, the potential experienced by $n_2$ decreases. 
This leads to the redistribution of $n_2$, i.e. it moves to the regions with 
weaker trapping potential. The redistribution will lead to $\delta P=P_1-P_2>0$
which decreases as one moves along the aforementioned curve. This will lead to
indentation of the inner interface. In our simulations, when the obstacle
is moved, along $x$-axis, from a point near the center towards the outer 
interface, indentation is suppressed. It confirms that this mechanism indeed 
aids the indentation of the interface. 


\subsection{Obstacle motion towards edge}
\begin{center}
\begin{figure}[ht]
\includegraphics[width=8.5cm] {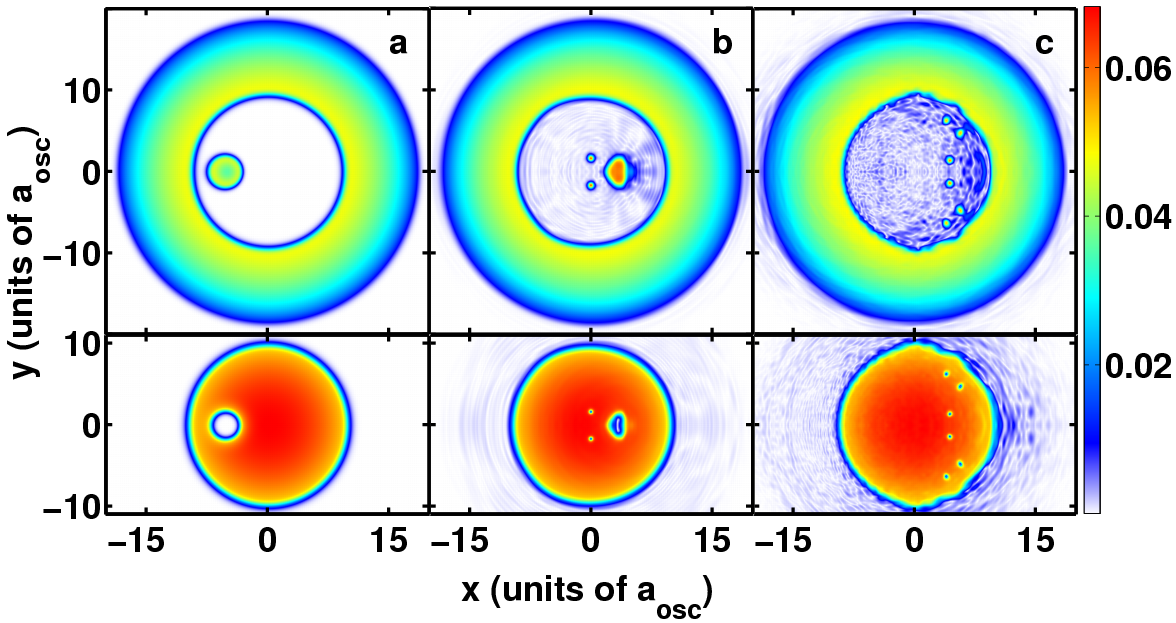}
\caption{The generation of triply charged vortex dipole in the
         $^{85}$Rb-$^{87}$Rb binary condensate with parameters {\em set a}.
         The obstacle potential, initially located at $x = -5.0 a_{\rm osc}$,
         is moved with a velocity of $220\mu$m/s up to $x = 5 a_{\rm osc}$.
         First, second, and third columns are the solutions at $t=0$s,
         $t=0.14$s and $t = 0.32$s, respectively.}
\label{triply_charged_vdipole}
\end{figure}
\end{center}

In this case, we start with the stationary state solution of 
$^{85}$Rb-$^{87}$Rb binary condensate for the parameters {\em set a} with 
obstacle laser potential maxima located at $x = -5.0a_{\rm osc}$. The static 
solution in this case is shown in first column of 
Fig.\ref{triply_charged_vdipole}, which has roughly doughnut shaped density 
distribution of the outer species along with non-zero density in the circle of 
influence of the obstacle.
 
\begin{center}
\begin{figure}[ht]
\begin{tabular}{cc}
\resizebox{42mm}{!}
{\includegraphics[trim = 0mm 0mm 0mm 0mm,clip, angle=0,width=8cm]
                 {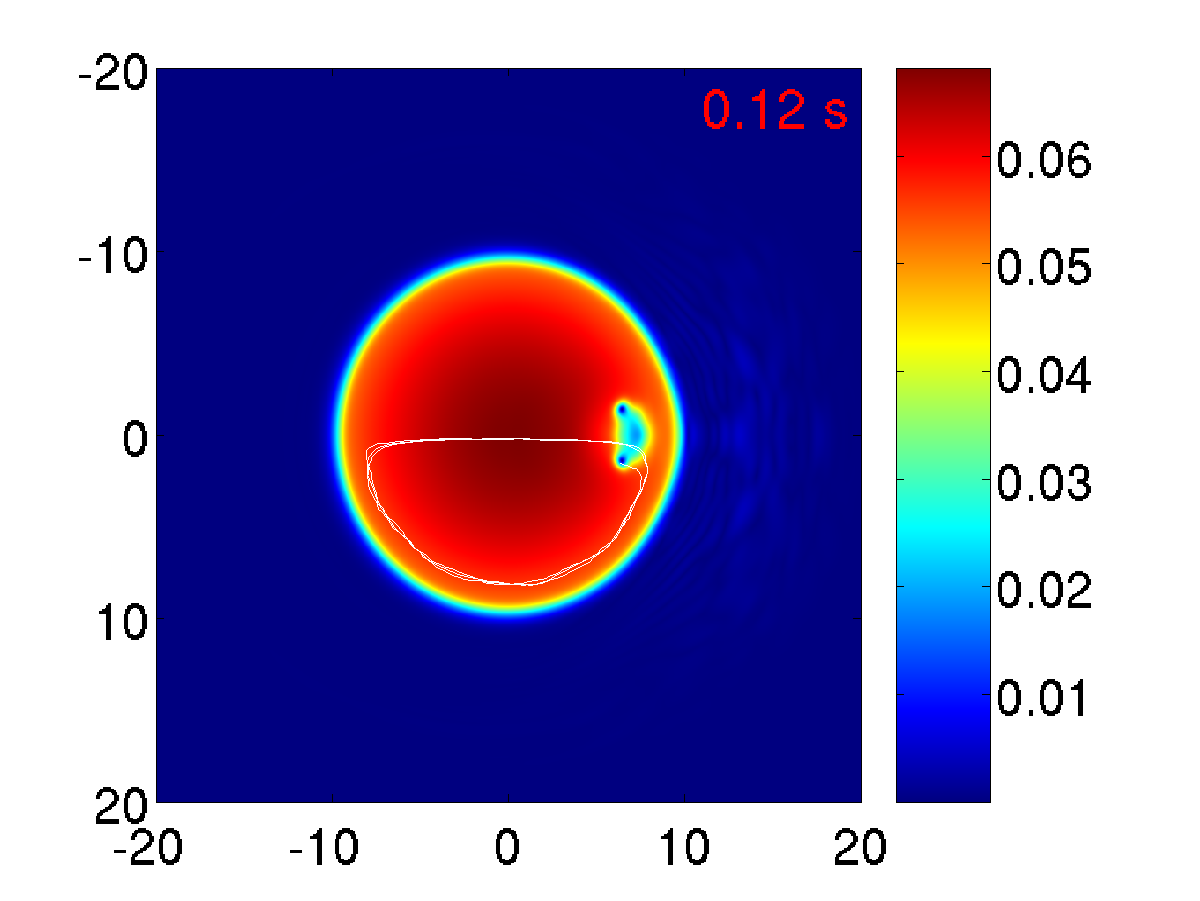}}&
\resizebox{42mm}{!}
{\includegraphics[trim = 0mm 0mm 0mm 0mm,clip, angle=0,width=8cm]
                 {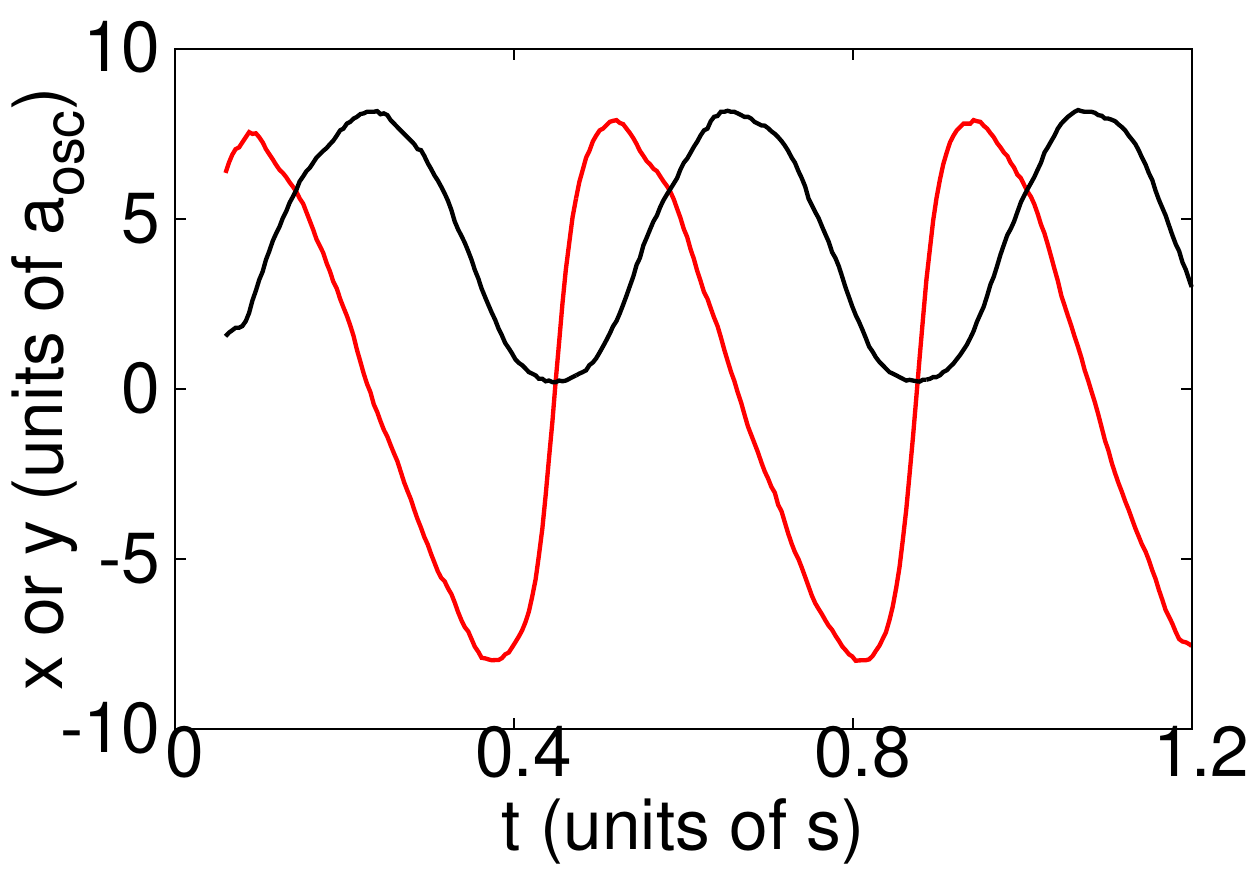}}\\
\end{tabular}
\caption{The figure on the left side shows the vortex dipole in 
$^{85}$Rb-$^{87}$Rb condensate with parameters {\em set a} at $t=0.12$s.
The white curve indicates the trajectory of lower vortex. The figure on
the right side shows the variation in position of the upper vortex with time.} 
\label{vdipole_trajectory}
\end{figure}
\end{center}

Such density distributions are prerequisite to create coreless vortices in the 
inner species. The obstacle is moved with a uniform speed of $220 \mu$m/s, 
progressively decreases in strength, and finally becomes zero at 
$x = 5.0a_{\rm osc}$. With this velocity, we observe the creation of triply 
charged coreless vortex dipole as is shown in last column of 
Fig.\ref{triply_charged_vdipole}. 
Singly and doubly charged vortex dipoles can be generated by moving the obstacle
potential at lower velocities. For example, 
Fig.\ref{vdipole_trajectory} shows the singly charged coreless vortex dipole 
and the trajectory followed by it for the $^{85}$Rb-$^{87}$Rb binary condensate 
with parameters {\em set a}. The obstacle laser potential moves from 
$x=0a_{\rm osc}$ at $t=0s$ to $x=8.0a_{\rm osc}$ with the speed of 
$200\mu$m/s.

\begin{center}
\begin{figure}[ht]
\includegraphics[width=8.5cm] {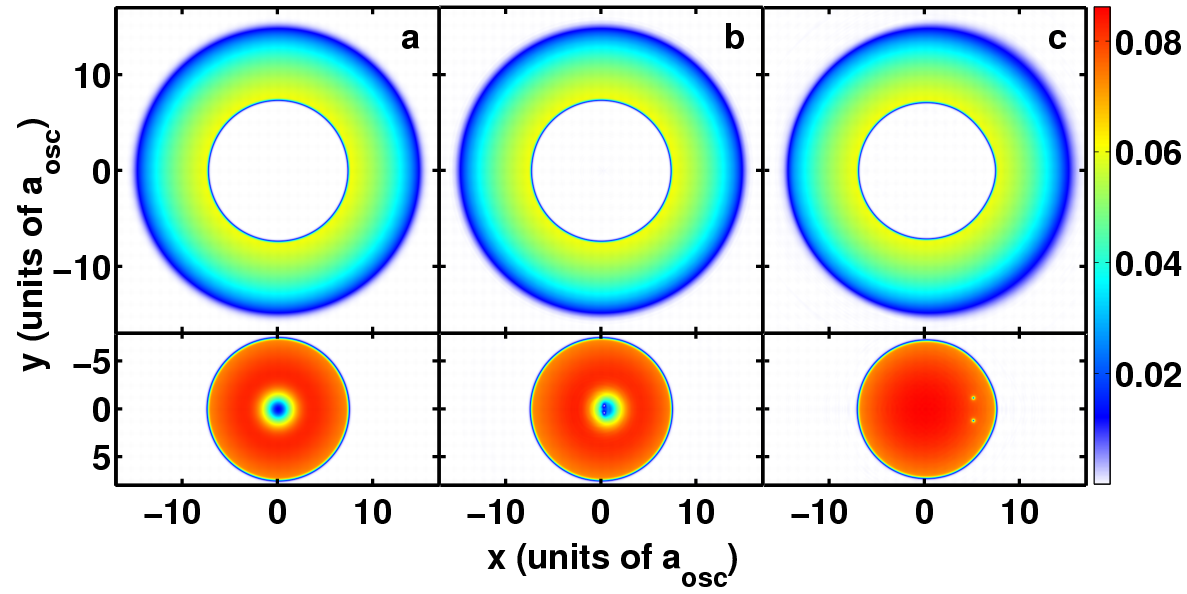}
\caption{The generation of normal vortex dipoles in the $^{85}$Rb-$^{87}$Rb
binary condensate with $a_{11} = 51a_0$, $N_1=N_2=10^6$, and the rest of the
parameters same as those defined in parameters {\em set a}. The obstacle
potential, initially located at $x = 0 a_{\rm osc}$, is moved with a velocity
of $200\mu$m/s up to $x = 5 a_{\rm osc}$. First, second, and third columns
are the absolute values of wave functions at $t=0$s, $t=0.01$s, and
$t = 0.14$s, respectively.}
\label{normal_vdipole}
\end{figure}
\end{center}

We also considered the $^{85}$Rb-$^{87}$Rb with $a_{11} = 51a_0$, 
$N_1 = N_2 = 10^6$, and the rest of the parameters same as those in parameter 
{\em set a}. The obstacle laser potential is initially located at origin 
$x = 0 a_{\rm osc}$. The stationary solution in this case is strongly phase
separated ($a_{12}\gg\sqrt{a_1a_2}$), and the circular region of influence of 
obstacle potential is devoid of atoms of outer component as is shown in first 
column of Fig.\ref{normal_vdipole}. In our simulation, the obstacle potential 
potential traverses a distance of $5a_{\rm osc}$ along x-axis with a speed 
of $200\mu$m/s. We observe that the vortex dipole generated in the inner 
component is a normal vortex dipole with empty core. This is
consistent with the TF results shown in Fig.~\ref{fig_coreless_vortex_dipole2}.


\section{Conclusions}
We have studied the motion of the Gaussian obstacle potential through a 
phase-separated binary condensate in pan-cake shaped traps. The motion of the 
obstacle leads to the generation of either coreless or normal vortex dipole 
depending upon the interaction parameters. The interaction parameters suitable
for coreless vortex dipole can also be employed to have obstacle assisted 
transport of one species across another. We have used both a variational 
{\em ansatz} based method, applicable when $N_ia_{ii}/a_{\rm osc}\sim 1$, and 
TF approximation based method, applicable when $N_ia_{ii}/a_{\rm osc}\gg 1$,
to analyze the energetic stability of coreless versus normal vortices.
Our studies show that using Feshback resonances to tune one of the scattering 
length, e.g. scattering length of $^{85}$Rb in $^{85}$Rb-$^{87}$Rb, it should
be experimentally possible to create coreless vortices in phase-separated
binary condensates.


\begin{acknowledgements}
We thank S. A. Silotri, B. K. Mani, and S. Chattopadhyay for very useful 
discussions. The numerical computations reported in the paper were done on
the 3 TFLOPs cluster at PRL. The work of PM forms a part of Department of 
Science and Technology (DST), Government of India sponsored research project.

\end{acknowledgements}


\end{document}